\documentclass[conference]{IEEEtran}
\IEEEoverridecommandlockouts
\usepackage{amsmath,amssymb,amsfonts}
\usepackage{mathtools}
\usepackage{graphicx}
\usepackage{textcomp}
\usepackage{xcolor}
\def\BibTeX{{\rm B\kern-.05em{\sc i\kern-.025em b}\kern-.08em
    T\kern-.1667em\lower.7ex\hbox{E}\kern-.125emX}}

\usepackage[style=ieee,backend=biber,bibencoding=utf8,sorting=none,maxbibnames=99]{biblatex}
\bibliography{bibliography}

\usepackage{algorithm}
\usepackage[noend]{algpseudocode}

\usepackage{multirow}
\usepackage{enumitem}

\begin{document}

\title{A Systematic Procedure for Topological Path Identification with Raw Data Transformation in Electrical Distribution Networks}

%%%%%   BY COLUMNS    %%%%%
% \author{\IEEEauthorblockN{Maurizio Vassallo}
% \IEEEauthorblockA{\textit{Electrical Engineering} \\
% \textit{University of Liège}\\
% Liège, Belgium \\
% mvassallo@uliege.be}
% \and

% \IEEEauthorblockN{Alireza Bahmanyar}
% \IEEEauthorblockA{\textit{Intelligent Systems Solutions} \\
% \textit{Haulogy}\\
% Neupré, Belgium \\
% alireza.bahmanyar@haulogy.net}
% \and

% \IEEEauthorblockN{Laurine Duchesne}
% \IEEEauthorblockA{\textit{Intelligent Systems Solutions} \\
% \textit{Haulogy}\\
% Neupré, Belgium \\
% laurine.duchesne@haulogy.net}
% \and

% \IEEEauthorblockN{Adrien Leerschool}
% \IEEEauthorblockA{\textit{Intelligent Systems Solutions} \\
% \textit{Haulogy}\\
% Neupré, Belgium \\
% adrien.leerschool@haulogy.net}
% \and

% % \IEEEauthorblockN{Simon Gerard}
% % \IEEEauthorblockA{\textit{\textcolor{red}{dept}. name of organization (of Aff.)} \\
% % \textit{Resa}\\
% % Liège, Belgium \\
% % simon.gerard@resa.be}
% \IEEEauthorblockN{Simon Gerard}
% \IEEEauthorblockA{\textit{Resa} \\
% Liège, Belgium \\
% simon.gerard@resa.be}
% \and

% \IEEEauthorblockN{Thomas Wehenkel}
% \IEEEauthorblockA{\textit{Resa} \\
% Liège, Belgium \\
% thomas.wehenkel@resa.be}
% \and

% \IEEEauthorblockN{Thomas Wehenkel}
% \IEEEauthorblockA{\textit{Resa} \\
% Liège, Belgium \\
% thomas.wehenkel@resa.be}
% \and

% \IEEEauthorblockN{Damien Ernst}
% \IEEEauthorblockA{\textit{LTCI, Télécom Paris} \\
% \textit{Institut Polytechnique de Paris} \\
% Paris, France \\
% dernst@uliege.be}
% }

\author{\IEEEauthorblockN{Maurizio Vassallo\IEEEauthorrefmark{1},
 Alireza Bahmanyar\IEEEauthorrefmark{2},
 Laurine Duchesne\IEEEauthorrefmark{2},
 Adrien Leerschool\IEEEauthorrefmark{2},\\
 Simon Gerard\IEEEauthorrefmark{3},
 Thomas Wehenkel\IEEEauthorrefmark{3},
 Damien Ernst\IEEEauthorrefmark{1}\IEEEauthorrefmark{4}
}

\IEEEauthorblockA{\IEEEauthorrefmark{1} Department of Electrical Engineering and Computer Science, Liège, Belgium}
\IEEEauthorblockA{\IEEEauthorrefmark{2} Intelligent Systems Solutions, Haulogy, Neupré, Belgium}
\IEEEauthorblockA{\IEEEauthorrefmark{3} RESA, Liège, Belgium}
\IEEEauthorblockA{\IEEEauthorrefmark{4} LTCI, Télécom Paris, Institut Polytechnique de Paris, France}
\IEEEauthorblockA{\{mvassallo, dernst\}@uliege.be}
\IEEEauthorblockA{\{alireza.bahmanyar, laurine.duchesne, adrien.leerschool\}@haulogy.net}
\IEEEauthorblockA{\{simon.gerard, thomas.wehenkel\}@resa.be}
}

\maketitle

\begin{abstract}
This paper introduces a systematic approach to address the topological path identification (TPI) problem in power distribution networks.
Our approach starts by listing the DSO's raw information coming from several sources.
The raw information undergoes a transformation process using a set of transformation functions. This process converts the raw information into well-defined information exploitable by an algorithm.
Then a set of hypothetical paths is generated, considering any potential connections between the elements of the power distribution system. This set of hypothetical paths is processed by the algorithm that identifies the hypothetical paths that are compatible with the well-defined information.
This procedure operates iteratively, adapting the set of transformation functions based on the result obtained: if the identified paths fail to meet the DSO's expectations, new data is collected, and/or the transformation functions found to be responsible for the discrepancies are modified.
The systematic procedure offers practical advantages for DSOs, including improved accuracy in path identification and high adaptability to diverse network configurations, even with incomplete or inaccurate data. Consequently, it emerges as a useful tool for the construction of digital twins of power distribution networks that aligns with DSO expectations.
\end{abstract}


\section{Introduction}
The distribution of electricity within power grids is a complex process characterised by networks of interconnected paths that transmit energy from a source to individual consumers. Within these intricate webs, each customer is generally associated with two types of paths: active paths, responsible for the distribution of electricity, and backup paths, which are used in case of emergencies. 
The identification of customer paths is important for distribution system operators (DSOs) for several reasons.
First, knowing the topological paths can serve as a vital tool to identify and inform customers about potential power outages due to network maintenance.
Second, it allows DSOs to proactively anticipate and address potential power quality issues within the network, such as overvoltages and outages.
Third, knowledge of backup paths can be useful to keep power supply, especially during emergencies.\\
% Fourth, the knowledge of the topological paths can be used for network planning studies, such as upgrading transformers or lines.\\

\noindent Solving the topological path identification (TPI) problem involves the identification of both active and backup paths for each customer in a given power distribution system.\\
DSOs try to solve the TPI problem, but unfortunately, they use case-specific techniques that might not always be practical in reality. For example, DSOs often rely on GPS data of the elements. They incorrectly assume this data to be accurate and complete. However, considering the scale of the networks and the limited precision of the GPS devices, there may be some missing values or inaccuracies in the GPS recordings that must be taken into account. Moreover, DSOs tend to concentrate only on a single active path for each customer, despite the potential existence of multiple paths that align with their information. Additionally, DSOs often overlook the significance of backup power supply paths, ignoring them. Therefore, DSOs lack a comprehensive framework, resulting in an inaccurate or incomplete identification of the paths, with consequent waterfall effects on other network operations.\\

\noindent Many studies have attempted to address the TPI problem based on the data available to DSOs. Among these techniques, some strategies use the concept of distance between elements to identify the paths. For example, in \cite{NavarroEspinosa2015RECONSTRUCTIONOL}, the authors propose an identification methodology that establishes connections between customers and the nearest feeder based on their spatial proximity.
% Such methodology is based on the breadth first search (BFS) algorithm, which visits every node of the power network to identify clusters and then connect them in order to have a network without interruptions.
Similarly, the authors in \cite{PCSSMGIS}, propose a work to identify the connectivity of elements of a low-voltage power network based on the distance of their coordinates, moreover, the authors identify and rectify connectivity errors between the elements that may arise in the network.
% such as incorrect connectivity and disconnected elements and rings.
In \cite{GDGMBPAD}, authors develop an algorithm to transform the data from the OpenStreetMap database into GPS information about the nodes to reconstruct the electrical power grid.
% This algorithm then analyses existing nodes, connecting junctions given a specified distance.
Alternative techniques take advantage of the smart meters' measurements.
In \cite{zepu_knowledge_2020}, the authors identify a low-voltage distribution network based on GIS data, energy market data, and network measurements.
% They begin by constructing a problem-specific ontology, listing various attributes, and then establishing relationships between these attributes.
The authors in \cite{LVNTIISMM} use the voltage and current time-series from a limited number of smart sensors to infer the network topology and the lines' impedance.
Similarly, the approach in \cite{IPDNTVCA} explores the correlation properties of voltage measurements to estimate the power distribution network topology. Authors in \cite{dejongh2023datadriven}, identify the switch states, phases, and line parameters of a low-voltage grid in German using the voltage and current measurements of a limited number of smart sensors.
% Given the results obtained in different fields, some other works propose machine learning (ML) solutions.
% The work \cite{fenrg.2020.613331} proposes a hybrid approach involving grid knowledge and a graph neural network in an iterative process where the neural network first proposes a graph, which is then  using knowledge inference.
% Authors in \cite{ADGWN} introduce an adaptive dual-channel graph wavelet neural network that identifies abnormal customer-transformer relationships by combining the network topology and customer consumption data through two interconnected graph wavelet networks.
% The work \cite{10.1371} proposes a strategy to reconstruct the network and to detect the changes in switching devices’ status using different ML techniques, such as k-nearest neighbours and support vector machine.
Despite the interesting results obtained, the prior research exhibits a dual set of limitations that tend to produce suboptimal solutions. First, they do not explicitly mention how each piece of raw information is transformed into well-defined information. This can be a problem if the output of their method is incorrect, since DSOs or data scientists may not easily identify the source of the error. Second, the mentioned works mainly focus on identifying a single active path among the many compatible ones, and they ignore the backup paths.
In other words, they solve the TPI problem in an unstructured and partial way.\\
\vspace{-4mm}
\vspace{1em}

\noindent To address the limitations of existing approaches, this paper presents a systematic procedure for solving the TPI problem. This systematic approach is characterised by listing an initial set of the DSO's raw information, coming from several sources. We then define a set of transformation functions that are used to transform the raw information into well-defined information exploitable by an algorithm.
% The raw information includes a diverse range of formats, including texts, images, and spreadsheets, which require some transformations to be effectively used.
We define a set of hypothetical paths, and we narrow it down by identifying the feasible paths that are compatible with the set of well-defined information.
The compatible paths are evaluated with a diagnostic function.
If these feasible paths fail to meet the evaluation criteria, new data is collected, and/or the transformation functions found to be responsible for the discrepancies are modified to ensure that the compatible paths align with the DSO's requirements. 
The systematic procedure of listing all the information and transformations can effectively complement existing methods, hence augmenting their capabilities and leading to more accurate results.
Moreover, the adaptability of this procedure to any available data, from GIS data to smart meter recordings, makes it suitable for every network and different DSO context.\\

% \noindent This paper presents a systematic and adaptive approach for the TPI problem of power distribution networks, specifically focusing on determining the active and backup paths between each customer and the different transformers.
% This work comes into existence to assist DSOs to manage the raw information coming from several sources.
% Our approach is based on the concept of functions that transform the raw information into well-defined data. and on the concept of hypothetical paths and the concept of feasible hypothetical paths that are paths compatible with the set of well-defined information.
% Given these settings, our proposed methodology is generic and adaptive to the amount of data available, enabling its application across various DSOs' contexts. Such methodology, indeed, can be applied to power distribution networks with different levels of knowledge, including only GIS data, smart meter data measurements and different kind of raw information.
% We argue that the key innovation of this paper lies in creating a seamless collaboration between network experts and data scientists for a more effective exploration of the networks' complexities.
% We prove our claim applying the proposed methodology to a real distribution network located in Belgium, demonstrating the effectiveness and applicability of the approach in a practical setting.\\

\noindent The rest of the paper is organised as follows. Section \ref{nettop} presents the definition of the power network's elements used in this work. Section \ref{tp} defines the concept of topological paths, with particular emphasis on real, active, backup, and hypothetical paths. 
% Moreover, a definition of similarity between paths is presented.
The problem statement is presented in Section \ref{ps}, outlining the objectives of the paper.
Section \ref{methodology} defines the methodology used to solve the problem. The methodology is applied to an academic example in Section \ref{accexample}.
Section \ref{discussion} presents complexity and scalability considerations of the proposed approach, presenting a potential alternative solution.
Section \ref{conclusion} concludes the work and discusses possible future works.

\section{Power network elements} \label{nettop}
% \noindent Given the set of elements, $\mathcal{E}$, it is possible to define a single topological path as a sequence of elements connected to each other. \\
Power distribution networks typically encompass a wide range of elements.
\noindent The set of elements $\mathcal{E}$ is defined as:
\begin{align}
    \mathcal{E} = \{e_1, e_2, ..., e_{|\mathcal{E}|}\}
\end{align}
\noindent where a single element inside the set $\mathcal{E}$ is denoted as $e_i$ and $|\mathcal{E}|$ represents the total number of elements in the set $\mathcal{E}$. \\

\noindent The set of attributes $\mathcal{A}$ is defined as follows:
\begin{align}
    \mathcal{A} = \{a_1, a_2, ..., a_{|\mathcal{A}|}\}
\end{align}
\noindent where a single attribute is denoted as $a_i$ and $|\mathcal{A}|$ represents the total number of attributes in the set $\mathcal{A}$.\\

\noindent A single element $e \in \mathcal{E}$ is defined as a list of attributes.
% Each element has at least 2 values: the coordinates and a type,
Formally:
\begin{align}
    e = (e.a_1, e.a_2, ..., e.a_{|e|})
\end{align}
where the dot symbol is used to access the attributes of the element $e$, $a_i$ is the $i$-th attribute, and $a_{|e|}$ is its last attribute.\\ %, with $a_i \in \mathcal{A}$. \\

\noindent The location attribute of the elements is identified by pairs of coordinates. 
Elements may have only one or more than one pair of coordinates.
Given one element, $e \in \mathcal{E}$, it is possible to access its coordinates as follows:
\begin{equation}
    c = e.coor
\end{equation}
where $c$ is the list of the element's coordinates. 
We refer to the $i$-th pair of the list of coordinates, $c$, of the element $e$ with the following notation: $e.coor_i$, with $1\leq i \leq |c|$, and $|c|$ the total number of pairs of coordinates of $c$. For example, $e.coor_1$ is the first pair of coordinates of the element, $e$, $e.coor_i$ is the $i$-th pair of coordinates, and $e.coor_{|c|}$ is the last one. \\

\noindent Each element within the network can be categorised based on its type attribute, serving as a characteristic that distinguishes it from other elements. The set of types, denoted as $\mathcal{T}$, includes all the possible types associated with the elements in $\mathcal{E}$:
\begin{align}
    \mathcal{T} = \{t_1, t_2, ..., t_{|\mathcal{T}|}\}.
\end{align}
\noindent Among the typical types, there may be $customer$ connection points, electrical $transformer$ stations, $switch$ devices, and other types that are possible to find in power distribution networks.\\
% \begin{align}
%      \mathcal{F} = \{ & customer, \\ \nonumber
%      &armoire,\\ \nonumber
%      &pole,\\ \nonumber
%      &tableau,\\ \nonumber
%      &transformer,\\ \nonumber
%      &switch \; open,\\ \nonumber
%      &switch \; close,\\ \nonumber
%      &underground,\\ \nonumber
%      &overhead \nonumber
%      \}
% \end{align}
% \noindent The function $Type()$ takes an element, $e \in \mathcal{E}$, as input and gives the type, $t \in \mathcal{T}$, of that element as output. This $Type()$ function is defined as follows:
% \begin{align}
%     t = Type(e).
% \end{align}
Given an element $e \in \mathcal{E}$, its type is accessed as follows:
\begin{align}
    t = e.type
\end{align}

% \noindent Each element in a distribution network has at least two attributes: geographic coordinates that define its location and a type that defines its characteristics. Some elements can have more than two attributes, such as elements of type $switch$, which have three attributes: the coordinates, a type, and a status representing whether the element of type $switch$ is closed (current can flow through it) or open (current cannot flow through it).\\

\noindent The $Subset()$ function is used to identify all elements of a specific type. This function takes two input parameters: the element set, $\mathcal{E}$, and a type $t \in \mathcal{T}$. The result of this function is a subset of $\mathcal{E}$, representing those elements of $\mathcal{E}$ associated with the specified type, $t$. Formally: 
\begin{align}
    Subset(\mathcal{E}, t) =  \{ e \in \mathcal{E} \; | \; e.type = t \}.
\end{align}
% Note that the output of the function $Subset()$ may be an empty set.\\

% \noindent Each element is characterised by a set of coordinates. Sometime, from some subsets ($customer$, $armoire$, $pole$, $tableau$, $transformer$, $switch \; open$ and $switch \; close$) this set is composed by only one value For some  other subsets ($underground$ and $overhead$) their elements are represented as a set of two elements: $|e| = 2, \forall e \in Subset(\mathcal{E}, underground)$. Similar to the elements in the path, the single elements of $e$ can be accessed as: $e^1$ and $e^2$. In this work, when we refer to an element, $e$, we refer to the set of its coordinates. \\

% For elements with more than one couple of coordinates, $|c|>1$, the first pair of coordinates, $c^1$, and the last one, $c^{|c|}$, represent the last pair of coordinates. \\
\noindent The distance between two elements is calculated using a $Dist()$ function. The $Dist()$ function takes two elements $e_i, e_j \in \mathcal{E}$ as input, and it outputs a scalar value, $d$, representing the minimum distance between their coordinates.
The $Dist()$ function is defined as follows:
\begin{align}
    % d &= Dist(c_i,c_j) = ||c_i-c_j||
    d &= Dist(e_i,e_j) = 
        \underset{
            \substack{\forall \;\! a \; \in \; e_i.coor, \\
            \forall \;\! b \; \in \; e_j.coor}
        }
        {\min} \; ||a-b||
\end{align}
%  The Dist() function is formally defined as follows:

%     If both inputs are elements (ei, ej ∈ E), the function calculates the minimum distance between all pairs of coordinates a ∈ ei.coor and b ∈ ej.coor using the Euclidean distance formula ||a − b||.

%     If one input is an element and the other is a coordinate, the function computes the distance between the element's coordinates and the specified coordinate.

% In equation form:
% d=Dist(ei,ej)=min⁡∀a∈ei.coor,∀b∈ej.coor∣∣a−b∣∣d=Dist(ei,ej)=min∀a∈ei.coor,∀b∈ej.coor ∣∣a−b∣∣
\noindent where $||\cdot||$ represents the Euclidean distance.
% Two elements $e_i, e_j \in \mathcal{E}$ are connected to another each other if their distance $Dist(e_i, e_j)$ is equal $0$.
\\
% In the case the distance between two elements is not allowed, the output, $d$, of $Dist()$ is $-1$.\\

\noindent Moreover, given an element, $e \in \mathcal{E}$, and a generic set of elements, $\mathcal{S}$, we define a function $Closest()$. This function identifies an element within the set $\mathcal{S}$ that is closest to the element $e$. The $Closest()$ function can be defined as follows:
\begin{align}
    s^* &= Closest(e,\mathcal{S}) = \underset{s \, \in \, S}{\arg\min} \bigl( Dist(e, s) \bigl)
\end{align}
where $s^*$ denotes the element of the set $\mathcal{S}$ that is the closest to $e$.
% Note that the generic set, $\mathcal{S}$, may be a subset of the set $\mathcal{E}$. In general, $\mathcal{S} \subseteq \mathcal{E}$.
% where $c^*$ denotes the pair of coordinates of the element $s$ from the set $\mathcal{S}$ that is closest to the element $e$.
% Note that the generic set, $\mathcal{S}$, may be a subset of the set $\mathcal{E}$. In general, $\mathcal{S} \subseteq \mathcal{E}$.

\section{Topological paths} \label{tp}
This section serves as the foundation for understanding the concepts of real, active, backup and hypothetical paths within the context of this work.

\subsection{Real paths} 
The flow of electricity in an electrical network generally follows a unique path from an MV/LV transformer to the customer. Between the customer and the transformer, there may be many other elements.
All of these connected elements together form a topological path. The set of all real paths, denoted as $\mathcal{P}$, is expressed as follows:
\begin{align}
    \mathcal{P} = \{p \in \mathcal{P}\; | \; &\text{$p$ is the real list of unique elements} \\ 
    &\text{from a customer to a transformers}\}. \nonumber
\end{align}
Each path, $p \in \mathcal{P}$, uniquely represents the real path that electricity can follow from an MV/LV transformer to the customer.
Due to the large scale of the power distribution networks and limitations of the recorded data, these paths are often not clearly known for the DSOs.\\

% provided that all the switches along this path are closed. \\
% and without any a priori information, paths can have any number of elements.
% For example, a path
% $p = (e_1, e_i, ..., e_j, e_9, e_{13})$
% means that the path $p$ is represented by the following elements: starting with an element $e_1$ (that is a customer, $Type(e_1) = customer$) connected to the node $n_i$ then some additional elements, the node $n_j$ connected to the node $n_9$ (that is a tableau, $Type(e_9) = tableau$) and finally the node $n_{13}$ (that is a transformer, $Type(e_{13}) = transformer$). \\
\noindent The $i$-th element of a path, $p \in \mathcal{P}$, is referred with the following notation: $p_i$, with $1\leq i \leq |p|$, and $|p|$ the total number of elements of the path $p$.
% For example, $p^1$ is always the customer,
% $p^2$ is the customer's next connection, and the final element,
% $p^{|p|}$, is the transformer.\\
For example, $p_1$ and $p_{|p|}$ are always a customer and the transformer, respectively.\\
Two elements in a path are considered connected if they directly follow each other. For a path $p$, $p_i$ and $p_j$ are connected if the condition $j=i+1$ is verified, with $i,j<|p|$.\\

\noindent Given an element of type $customer$, $e \in Subset(\mathcal{E}, customer)$, the set of paths $\mathcal{P}_{e}$, represents the subset of all paths in which the customer, $e$, is present as starting point. Formally, the subset $\mathcal{P}_{e}$ is determined using the following notation:
\begin{align}
    % \mathcal{P}_{e} = \{ p \in Subset(\mathcal{P}, customer) \; | \; p^{1}=e\}
    \mathcal{P}_{e} = \{ p \in \mathcal{P} \; | \; p_{1}=e\}.
\end{align}

\noindent The length of a given path $p \in \mathcal{P}$ is calculated with the following formula:
\begin{align}
    % l = LengthPath(p) = \sum_{i=1}^{|p|-1} Dist(p_j, p_{j+1})
    l & = LengthPath(p) \\
    & = \sum_{i=1}^{|p|-1} \underbrace{Dist(p_i,p_{i+1})}_{i}
    +
    \underbrace{\sum_{j=1}^{|p_i.coor|-1} Dist(p_i.coor_j, p_i.coor_{j+1})}_{ii}. \nonumber
\end{align}
The formula calculates the total length ($l$) of a given path ($p$) by summing the distances between consecutive elements ($i$) along the path and adding the length of a single element ($ii$). The length of a single element is determined by the sum of the distances between pairs of consecutive coordinates.

\subsection{Active and backup paths}
A single customer can have many paths to different transformers, but generally, only one path is active, while all the others are backup paths used in case of emergency. The status attribute of the switches defines the activity or inactivity of a path.\\

\noindent The set of active paths, $\mathcal{P}_a$ is defined as follows:
\begin{align}
        % \mathcal{P}_a = \{ p \in \mathcal{P} \; | \; \text{there are no switches or all switches in $p$ are close} \}
        \mathcal{P}_a = \{ p \in \mathcal{P} \; | \; &\text{$p$ contains no elements of} \\
        & \text{type switch with status open} \}. \nonumber
\end{align}
\noindent Similarly, the set of backup paths, $\mathcal{P}_b$, is formulated as follows:
\begin{align}
        % \mathcal{P}_i = \{ p \in \mathcal{P} \; | \; \text{there is at least one switch and at least one switch in $p$ is open} \}
        \mathcal{P}_b = \{ p \in \mathcal{P} \; | \; &\text{$p$ contains at least on element of} \\
        & \text{type switch with status open} \}. \nonumber
\end{align}

\noindent Given the set of paths, $\mathcal{P}_a$ and $\mathcal{P}_b$, the DSOs can effectively analyse the paths for each customer and identify both its active and backup paths. In particular, given an element of type customer, $e \in Subset(\mathcal{E},  customer)$, the sets of the customer's active and backup paths are identified as follows:
\begin{align}
    P_{a,e} & = \{ p \in \mathcal{P}_a \; | \; p_1=e \},\\
    P_{b,e} & = \{ p \in \mathcal{P}_b \; | \; p_1=e \}.
\end{align}
\subsection{Hypothetical paths}
In this context, hypothetical paths refer to a set of potential paths within the distribution network.
The term “hypothetical” emphasises the uncertainty in determining the topological paths due to incomplete or inaccurate information. Indeed, the hypothetical paths are constructed under the consideration that only the initial point of the paths is known, while all connections among the in-between elements to a transformer may be unknown.\\

% \noindent We define the set of hypothetical path, denoted as $\mathcal{H}$, as the collection of all the paths that could exist without having any information about the network, other than the set of elements $\mathcal{E}$ and the types of the different elements, $\mathcal{T}$:
\noindent The set of hypothetical paths, denoted as $\mathcal{H}$, is expressed as follows:
\begin{align}
    \mathcal{H} = \{h \in \mathcal{H} \; | \; & \text{$h$ is any list of unique elements} \\
    &\text{from a customer to a transformer}\}. \nonumber
\end{align}
% \begin{align}
%     \mathcal{H} = \{p \; | \; & \text{$p$ is any list of unique elements} \\ 
%     &\text{from a customer to a transformer} \nonumber \\
%     &\text{not knowing the next connections}\}. \nonumber
% \end{align}

% \begin{figure}[h]
%     \centering
%     \includegraphics[width=0.22\textwidth]{Images/H/ResaExample-Page-4A.drawio.png}
%     \includegraphics[width=0.22\textwidth]{Images/H/ResaExample-Page-4B.drawio.png}
%     % \includegraphics[width=0.22\textwidth]{Images/H/ResaExample-Page-4C.drawio.png}
%     % \includegraphics[width=0.22\textwidth]{Images/H/ResaExample-Page-4D.drawio.png}
    
%     \caption{Example of some possible situations of hypothetical path sets from a customer ($e_1$) to a transformer ($e_5$) given some generic elements ($e_2$ to $e_4$). Situation \textit{a)} represent no knowledge for the next connection of $e_1$ this leads to a large hypothetical path set, given by any combination of the colored dotted lines. Situation \textit{b)} presents some knowledge about the connection, so the hypothetical path set is given by only two elements: from customer $e_1$ to element $e_3$ following the colored solid connections and then the paths can finish directly with the sustation $e_5$ (blue dotted connection) or passing through the element $e_4$ first (blue and light blue connections).}
%     \label{fig:hp}
% \end{figure}

\newcommand*{\Perm}[2]{{}^{#1}\!P_{#2}}%

\noindent The size of the hypothetical path set denoted as $|\mathcal{H}|$, given a set of elements $\mathcal{E}$ is calculated with the following parameters.
\begin{itemize}
    \item $\mathtt{E}$: the total number of elements in the set $\mathcal{E}$, $\mathtt{E}=|\mathcal{E}|$.
    
    \item $\mathtt{C}$: the total number of elements of type $customer$ in the set $\mathcal{E}$, $\mathtt{C}=|Subset(\mathcal{E}, customer)|$.
    
    \item $\mathtt{T}$: the total number of elements of type $transformer$ in the set $\mathcal{E}$, $\mathtt{T}=|Subset(\mathcal{E}, transformer)|$.
    
    \item $\mathtt{R}=\mathtt{E}-\mathtt{C}-\mathtt{T}$: the remaining number of elements that can be connected to each other, excluding elements of type $customer$ and $transformer$.
    
    \item $\Perm{n}{k}$: the permutation formula, which calculates the number of ways to arrange $k$ items from a set of $n$ items ($\Perm{n}{k} = \frac{n!}{(n-k)!}$).
\end{itemize}
The following equation calculates the total number of possible hypothetical paths in the network:
\begin{align} \label{eq:h}
    |\mathcal{H}| = \sum_{i=0}^{|\mathtt{R}|} |\mathtt{C}| \cdot \Perm{|\mathtt{R}|}{i} \cdot |\mathtt{T}|.
\end{align}
% The Eq. (\ref{eq:h}) uses the function $\Perm{n}{k}$, which represents the permutation formula to calculate the number of ways to arrange $k$ items from a set of $n$ items.\\
Equation (\ref{eq:h}) can be interpreted as follows.
It starts by considering the simplest paths, which involve directly connecting a customer to a transformer in a single step ($i=0$). \\
Gradually increasing the number of steps, it considers longer paths that involve more intermediate elements. The elements are considered in various permutations ($\Perm{\mathtt{R}}{i}$).\\
The calculation ends when the maximum step ($i=R$) is reached, indicating the longest paths possible in the network.
\\
% The initial path length considered is $2$ ($\mathtt{p}$$\,=\,$$2$), representing a direct association between a customer and a transformer. Subsequently, paths of length $3$ are analysed, incorporating an element distinct from both customer and transformer types. This process iterates over progressively increasing path lengths, stopping at the maximum step defined as: $\mathtt{p}_{max}=\mathtt{E}-\mathtt{C}-\mathtt{T}$. \\

\noindent It is worth noting that the same notation used for the real path set can be extended to the hypothetical path set. For instance, when identifying the hypothetical paths associated with a specific customer, denoted as $e \in Subset(\mathcal{E},  customer)$, they can be expressed as follows:
% \vspace{-2mm}
\begin{align}
    \mathcal{H}_e = \{ h \in \mathcal{H} \; | \; h_1=e \}.
\end{align}

\section{Problem statement} \label{ps}
% The aim of DSOs is to determine a subset of the hypothetical paths $\mathcal{H}$, based on a set of information $\mathcal{I}$, an approximation of the set of elements $\mathcal{E}$, given by the elements available to the DSO, and their types $\mathcal{T}$.
DSOs aim to identify a subset of the hypothetical paths set $\mathcal{H}$, that represent the closest approximation of the set of real paths $\mathcal{P}$, denoted as $\hat{\mathcal{P}}$. This identification process depends on a set of raw information, $\mathcal{I}$. Based on this information, it is possible to construct an approximate representation of the set of elements known to the DSO and their associated attributes, respectively denoted by $\hat{\mathcal{E}}$ and $\hat{\mathcal{A}}$.\\
% , and a set of rules used to identify the hypothetical paths that are compatible with the given information.
Ideally, we would compare the set of real paths, $\mathcal{P}$, and its approximation, $\hat{\mathcal{P}}$, using a rigorous metric to measure the similarity between the two sets. This metric would evaluate the degree of similarity between the sets, resulting in a similarity score ranging from $0$ to $1$. The evaluation of the two sets can result in one of the three possible outcomes: a perfect match indicated by a score of $1$, complete dissimilarity corresponds to a score of $0$, and partial congruence represented by any score between $0$ and $1$.\\
% \begin{itemize}
%     \item They are congruent and the similarity score between them is $1$.
%     \item They are totally different and the similarity score between them is $0$
%     \item The subset of hypothetical paths is between being congruent and totally different from $\mathcal{P}$ and the similarity between them is a value between $(0,1)$.
% \end{itemize}
Unfortunately, this evaluation is not possible since the set of real paths, $\mathcal{P}$, is generally unknown to DSOs. For solving the lack of a proper evaluation between sets, the solution found by our procedure is validated with a $Diagnostic()$ function. This function serves as a critical tool for assessing the validity of the solution. The $Diagnostic()$ function evaluates the found paths by checking any possible discrepancy from reality.
% With the successful identification of hypothetical paths, DSOs can also be interested in the sets of active and inactive paths. \\

\section{Methodology} \label{methodology}
This section outlines the systematic procedure proposed in this study to address the TPI problem in power distribution networks. 
The main goal of this methodology is to identify the hypothetical paths in $\mathcal{H}$ that are compatible with a set of well-defined information. This set represents the closest approximation of the set of real paths $\mathcal{P}$, denoted as $\hat{\mathcal{P}}$.

\subsection{Raw information, transformation functions and well-defined information}
Raw information refers to available information or data that comes from various sources. Usually, the raw information is provided in an unorganised way, such as text, images, and spreadsheets, that cannot be potentially exploited directly.
The set of raw information is denoted as follows: $\mathcal{I} = \{i_1, i_2, ..., i_{|\mathcal{I}|} \}$.\\
% For example, an information, $i_j \in \mathcal{I}$ may state that DSOs \textit{want to minimise the length of the lines connecting customers to the feeder}.\\

\noindent Transformation functions are responsible for transforming the available raw information into clear knowledge that is relevant within the specific context of a power network.
The set of transformation functions is denoted as follows: $\mathcal{F} =\{ f_1, f_2, ..., f_{|\mathcal{F}|} \}$. Mathematically, the transformation functions take one or more pieces of raw information as input and return the transformed information as output. \\ 

\noindent Well-defined information refers to the transformation of the raw information set $\mathcal{I}$, using the transformation function set $\mathcal{F}$. This well-defined information forms the basis to provide a structured foundation, essential for identifying compatible paths.
The set of well-defined information is denoted as follows: $\mathcal{I}' =\{ i'_1, i'_2, ..., i'_{|\mathcal{I}'|} \}$, with a single piece of well-defined information being the output of a transformation functions, i.e. $i' = f(i)$. \\

\noindent Given the set of well-defined information, $\mathcal{I}'$, the goal is to identify a subset of the hypothetical path set compatible with $\mathcal{I}'$, denoted as $\mathcal{H}^{\mathcal{I}'}$.
This set represents the best approximation of the set of real paths, denoted as $\hat{\mathcal{P}}$, and it is considered the solution for the TPI problem.
% The set $\hat{\mathcal{H}}^{\mathcal{I}'}$ is only an approximation because in large cases the set of hypothetical paths $\mathcal{H}^{\mathcal{I}'}$ may be impossible to find.

\subsection{Methodology steps}
\noindent Such methodology is performed in six steps:
\begin{enumerate}
    \item The first step involves listing the information available to the DSO, denoted as $\mathcal{I}$. 
    
    \item Then establish an initial set of transformation functions, denoted as $\mathcal{F}$. 
    
    \item Extract the relevant well-defined information, denoted as $\mathcal{I}'$, from the raw available data $\mathcal{I}$, using the transformation functions $\mathcal{F}$. \label{ms:2}
    
    \item With the well-defined information $\mathcal{I}'$, use an algorithm to identify a subset of the hypothetical paths compatible with the set of well-defined information, $\mathcal{I}'$, denoted as $\mathcal{H}^{\mathcal{I}'}$.
    
    \item Evaluate the solution found with the $Diagnostic()$ function.
    \begin{itemize}
        \item If any issue is detected, new data is collected, and/or the transformation functions are modified. Return to step \ref{ms:2}. 
        \item Otherwise, if no issue is observed, the set $\mathcal{H}^{\mathcal{I}'}$ is considered as the solution for the problem, and it represents the closest representation of the set of real paths $\mathcal{P}$, denoted as $\hat{\mathcal{P}}$. Continue to step \ref{ms:6}.
    \end{itemize}
    
    \item Once the solution $\hat{\mathcal{P}}$ is found, identify the customers' active and backup paths. \label{ms:6}
\end{enumerate}
% Essentially, this methodology follows an iterative process that begins by defining certain functions that are used to transform and alter the raw information into well-defined information. The well-defined information is then used to determine an approximation of the feasible hypothetical path set $\hat{H}^{\mathcal{I}'}$. Following this, an evaluation of the obtained solution is performed using a $Diagnostic()$ function. If any issues or discrepancies are detected, some adjustments are made to the transformation functions.
% This process iterates until the subset of hypothetical paths, $\hat{H}^{\mathcal{I}'}$, meets the expectations of the DSO.
These six simple steps collectively constitute a systematic and adaptive procedure that enables the identification of both active and backup paths, ensuring compatibility with the well-defined information.

\section{Academic example}
 \label{accexample}
To illustrate the methodology used to solve the TPI problem, an academic example of a power distribution network with a limited number of elements is considered.
This example represents a network where two main feeders are connected to two MV/LV transformers, supplying some customers through active and backup paths.\\
\noindent The example, along with the associated data and code, is publicly available on the GitHub repository \textit{\textcolor{black}{https://github.com/TPIproblem/TPIpaper}}, featuring an open-source Apache v2 licence.

\subsection{Real network}
\begin{figure}[h]
    \centering
\includegraphics[width=0.42\textwidth]{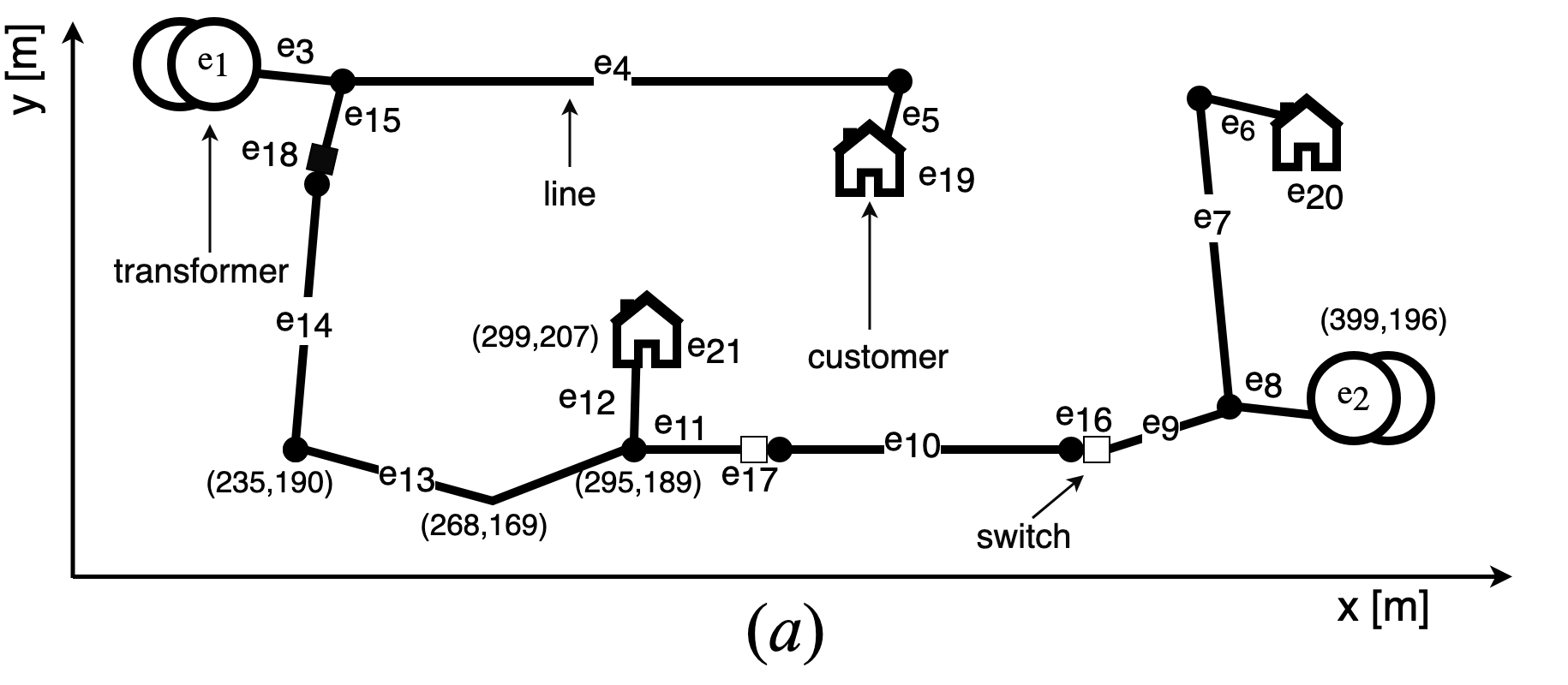}
\includegraphics[width=0.42\textwidth]{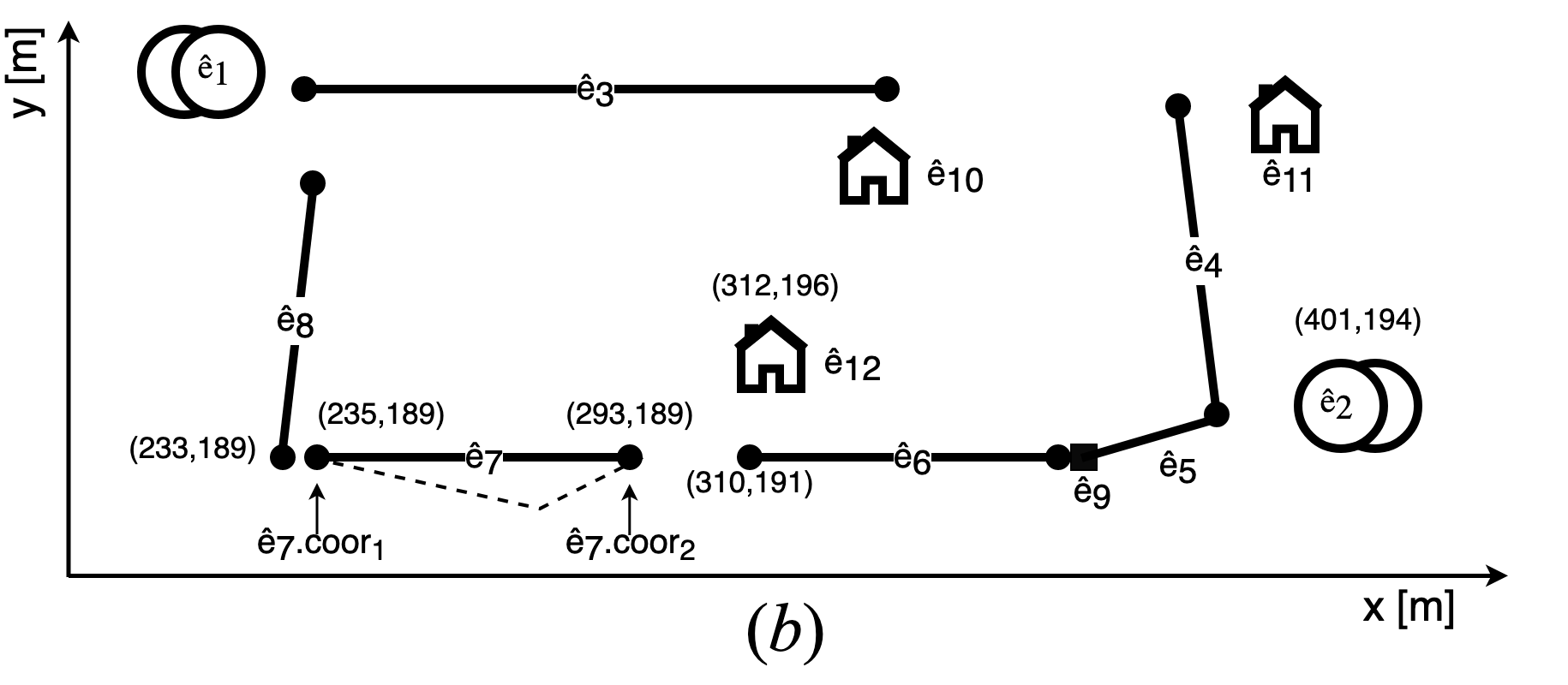}
    
\caption{Real network, Fig. \ref{fig:realrec}a, and the approximate network when considering only the elements known by the DSO, Fig. \ref{fig:realrec}b. It is possible to note some inaccuracies in the GPS location and some missing elements.}
    \label{fig:realrec}
\end{figure}

The real configuration of the network under consideration is illustrated in Fig. \ref{fig:realrec}a. It includes different kinds of elements such as transformers, lines, switches, and customers. Each element's position is defined by one or more pairs of coordinates that localise them within a 2-D plane, as it is possible to see for some representative elements: $e_2$, $e_{13}$ and $e_{21}$. Switch elements are visually differentiated by an empty square or a filled square, to indicate their status: open or closed, respectively. More details about the coordinates of the elements can be found in the Appendix.
% Generally, DSO do not have such accurate data to represent the grid.

% In such case, there are different elements, like: customers, transformers, lines and switches as shown in Fig. \ref{fig:example}a.
% The example is for simplicity limited to one customer, but the same considerations can be extended to a case with many customers.
\subsection{Raw information}
We assume the set of raw information available to the DSO, denoted as $\mathcal{I}$, is provided by some pieces of information:
% For instance, some raw information may be:
\begin{enumerate}
    \item[$i_1$:] Documents containing the DSO's list of elements, their coordinates, and their types. However, some elements are missing, and some coordinates are inaccurate, as illustrated in Fig. \ref{fig:realrec}b.
    \item[$i_2$:] The DSO lacks information about the customers' connections to the network, but their general practice is to minimise the cost of connecting customers to the grid. 
    \item[$i_3$:] Information collected from a technical meeting with the DSO's experts where they stated that, for economic considerations (cost per meter) and efficiency concerns (energy loss per meter), the connections between two elements should not be too long.
    % \item[$i_3$:] Information regarding the radial configuration of the distribution network.
    \item[$i_4$:] DSO's general practice is to reduce energy losses by minimising the total path length of each customer.
    \item[$i_5$:] The knowledge that each customer is supplied by exactly one transformer.
\end{enumerate}

\subsection{Well-defined information}
To identify accurate paths, our methodology uses a set of transformation functions, denoted as $\mathcal{F}$. These transformation functions convert the set of information available to the DSO, denoted as $\mathcal{I}$, into well-defined information, denoted as $\mathcal{I}'$.
This well-defined information is used by an algorithm to find an approximation of the set of real paths $\mathcal{P}$, expressed as $\hat{\mathcal{P}}$. 
\\

\begin{figure}[h!]
    \centering
\includegraphics[width=0.35\textwidth]{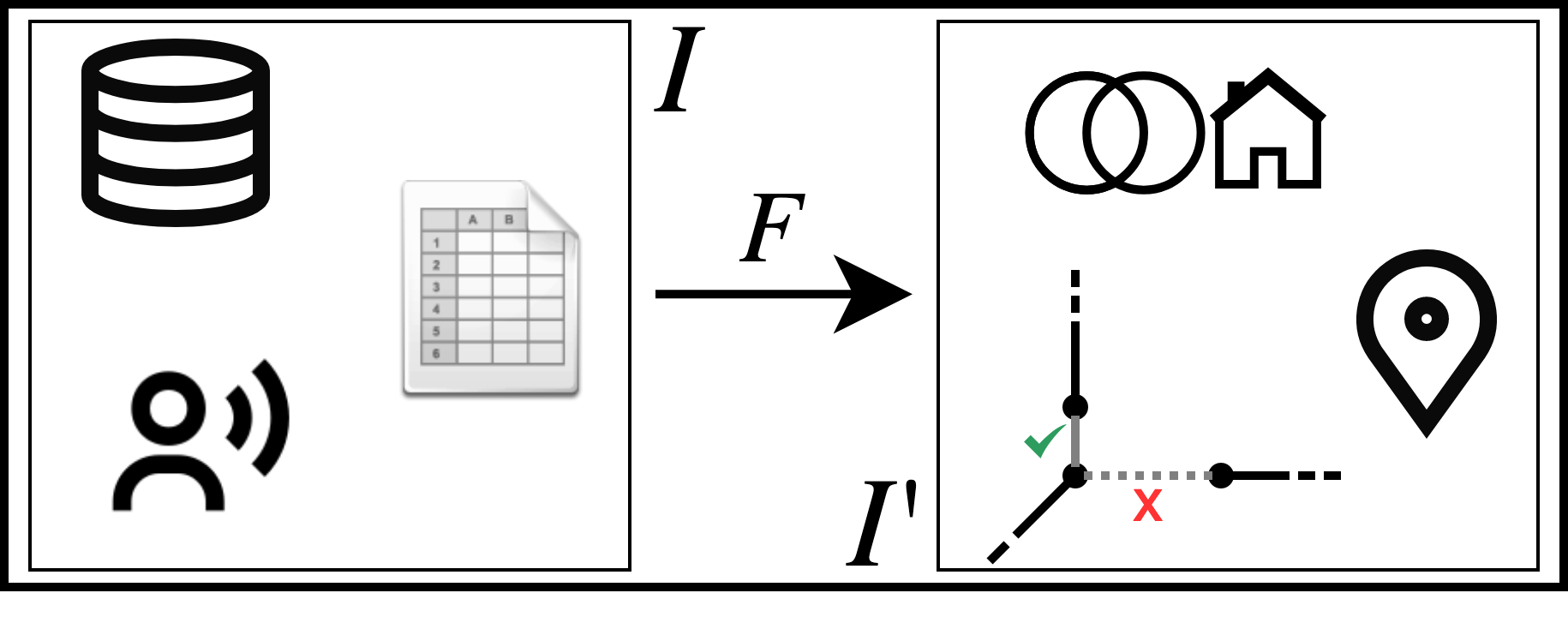}
    
\caption{Transformation functions transforming different kinds of data into well-defined information.}
    \label{fig:tf}
\end{figure}

\noindent For example, the transformation functions transform the raw information, given by database recordings, spreadsheet files, or technical meetings, into well-defined information, like power distribution network elements and their type, GPS coordinates, and business rules, as shown in Fig. \ref{fig:tf}.

\noindent In this case, the well-defined information $\mathcal{I}'$, obtained as a result of transforming the raw information $\mathcal{I}$ with the transformation functions $\mathcal{F}$, is presented as follows.
\begin{enumerate}
    \item[$i'_1$]$= f_1(i_1)$: set of elements, $\hat{\mathcal{E}}$, and their attributes, $\hat{\mathcal{A}}$.
    
    \item[$i'_2$]$= f_2(i_2)$: elements of type $customer$ are connected to the closest element of type $line$. \\
    Therefore, for every path $h \in \mathcal{H}^{\mathcal{I}'}$, the condition $h_2 = Closest\big( Subset(\hat{\mathcal{E}}, line), h_1 \big)$ must be satisfied.
    
    \item[$i'_3$]$= f_3(i_3)$: elements of type $transformer$, $line$, and $switch$ are connected to the closest element only if the distance calculated using the elements GIS coordinates is smaller than a maximum value $R$. \\
    Therefore, for every path $h \in \mathcal{H}^{\mathcal{I}'}$ and for any $i<|h|-1$ the condition $h_{i+1} = Closest(\mathcal{E}, h_i)\text{ AND }Dist(h_i, h_{i+1}) < R$ must be satisfied.
    
    \item[$i'_4$]$ = f_4(i_4)$: the total length of a path, defined as the sum of the lengths of the single lines, is at most $L$. \\
    Therefore, for every path $h \in \mathcal{H}^{\mathcal{I}'}$ the condition $LengthPath(h) < L$ must be satisfied.
    
    % \item[$i'_5$]$ = f_5(i_5)$: the number of active paths for each element $e$ of type $customer$ is equal to one and only one. \\    Therefore, $|\mathcal{H}^{\mathcal{I}'}_{a,e}|=1$ must be satisfied for any $e \in Subset(\mathcal{E}, customer)$.
\end{enumerate}
% The objective of these well-defined information is to pose some rules to identify the paths that are compatible with the information available to the DSO.
Given the well-defined information $i'_1$, it is possible to define the set of elements known to the DSO, $\hat{\mathcal{E}}$, and their attributes $\hat{\mathcal{A}}$. Generally $\hat{\mathcal{E}}$ is a subset of $\mathcal{E}$ as there might be additional elements in the actual network that are not present in the data available to the DSO.\\

\noindent In such an academic example, the set $\hat{\mathcal{E}}$ is given by the elements $\hat{e}_1$ to $\hat{e}_{12}$ as shown in Fig. \ref{fig:realrec}b. The set of types, $\hat{\mathcal{T}}$, is composed of: $transformer$ (elements $\hat{e}_1$ and $\hat{e}_2$), $line$ ($\hat{e}_3$ to $\hat{e}_8$), $switch$ ($\hat{e}_9$) and $customer$ ($\hat{e}_{10}$ to $\hat{e}_{12}$). \\
\noindent Figure \ref{fig:realrec}b illustrates the geographical position of the elements in the network, as defined by $i'_1$.
In particular, it is possible to note that the position of the elements of type $transformer$ and $customer$ are identified by one pair of coordinates; elements of type $line$ are identified by two pair of coordinates; elements of type $switch$ are also identified by a single pair of coordinates, and they are represented differently, either as empty or filled, to better visualise their status attribute. An empty representation corresponds to a switch whose status is open, while a filled representation corresponds to a switch whose status is closed. \\
Table \ref{tab:elemprop} summarises the attributes of each element.\\
\begin{table}[h!]
\centering
\caption{Elements and their attributes}
\label{tab:elemprop}
\resizebox{\columnwidth}{!}{%
\begin{tabular}{|c|ccc|}
\hline
\multirow{2}{*}{\textbf{Element}} & \multicolumn{3}{c|}{\textbf{Attributes}}                                      \\ \cline{2-4} 
                     & \multicolumn{1}{c|}{\textbf{Coordinates}} & \multicolumn{1}{c|}{\textbf{Type}} & \textbf{Status}  \\ \hline
$\hat{e}_1$ and $\hat{e}_2$      & \multicolumn{1}{c|}{One pair}             & \multicolumn{1}{c|}{$transformer$} & - \\ \hline
$\hat{e}_3$ to $\hat{e}_8$       & \multicolumn{1}{c|}{Two}    & \multicolumn{1}{c|}{$line$}        & - \\ \hline
$\hat{e}_9$                & \multicolumn{1}{c|}{One pair}             & \multicolumn{1}{c|}{$switch$}      & Open or close \\ \hline
$\hat{e}_{10}$ to $\hat{e}_{12}$ & \multicolumn{1}{c|}{One pair}             & \multicolumn{1}{c|}{$customer$}    & - \\ \hline
\end{tabular}%
}
\end{table}

\noindent As mentioned, the data may not always be accurate or there may be some wrong or missing coordinates. For example, the customer $\hat{e}_{12}$, in the data available to DSO is closer to the element $\hat{e}_6$ while in reality, the customer is closer to the element $e_{13}$ (corresponding to the element $\hat{e}_7$). Furthermore, the element $\hat{e}_7$ is missing a pair of coordinates in the information available to the DSO (highlighted in the dashed line in Fig. \ref{fig:realrec}b). Indeed, for lines, DSOs have only the endpoints (illustrated as $\hat{e}_7.coor_1$ and $\hat{e}_7.coor_2$), while intermediate points may not be available. Therefore, in the reconstructed network, the line is modeled as a straight line between its two endpoints. In such a case, this approximation introduces the possibility that the length of the reconstructed line may deviate from the actual length in the real network.
% Given these data inaccuracies and missing values, the objective of the DSO is to determine an approximation of the paths that closely align with the actual network conditions.

\begin{figure}[h]
    \centering
\includegraphics[width=0.45\textwidth]{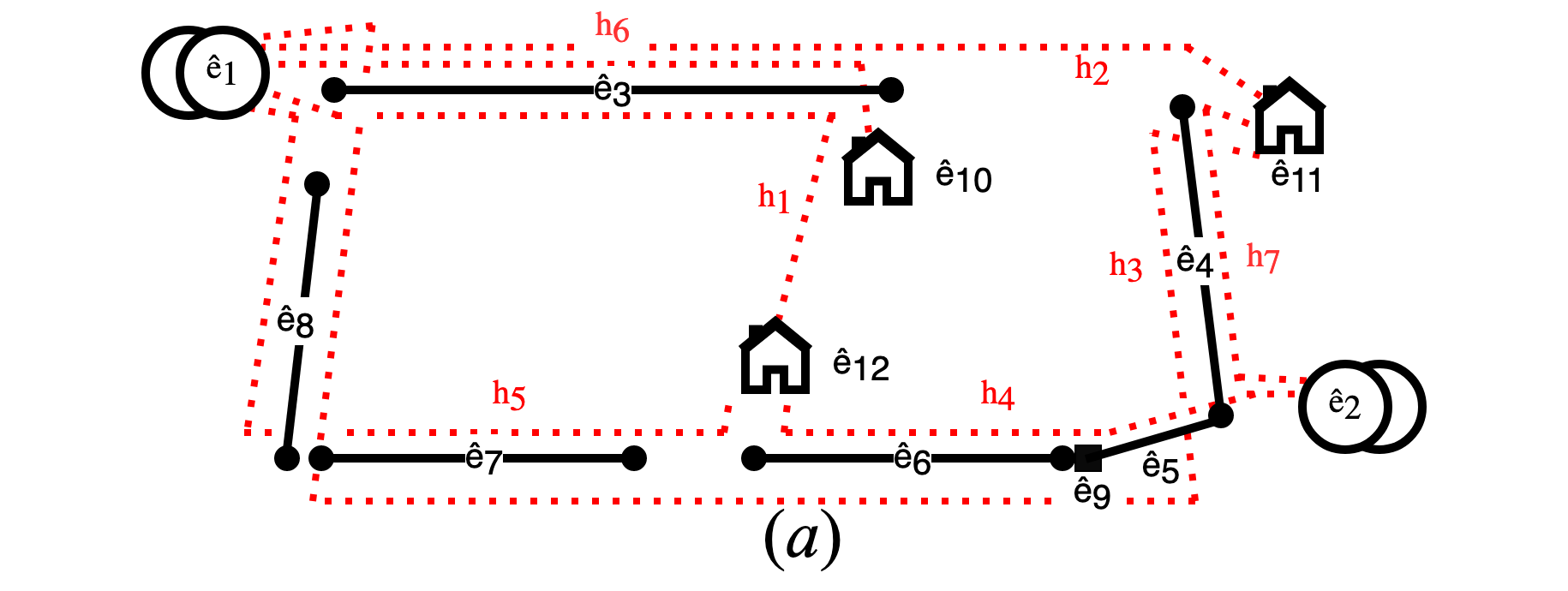}
\includegraphics[width=0.45\textwidth]{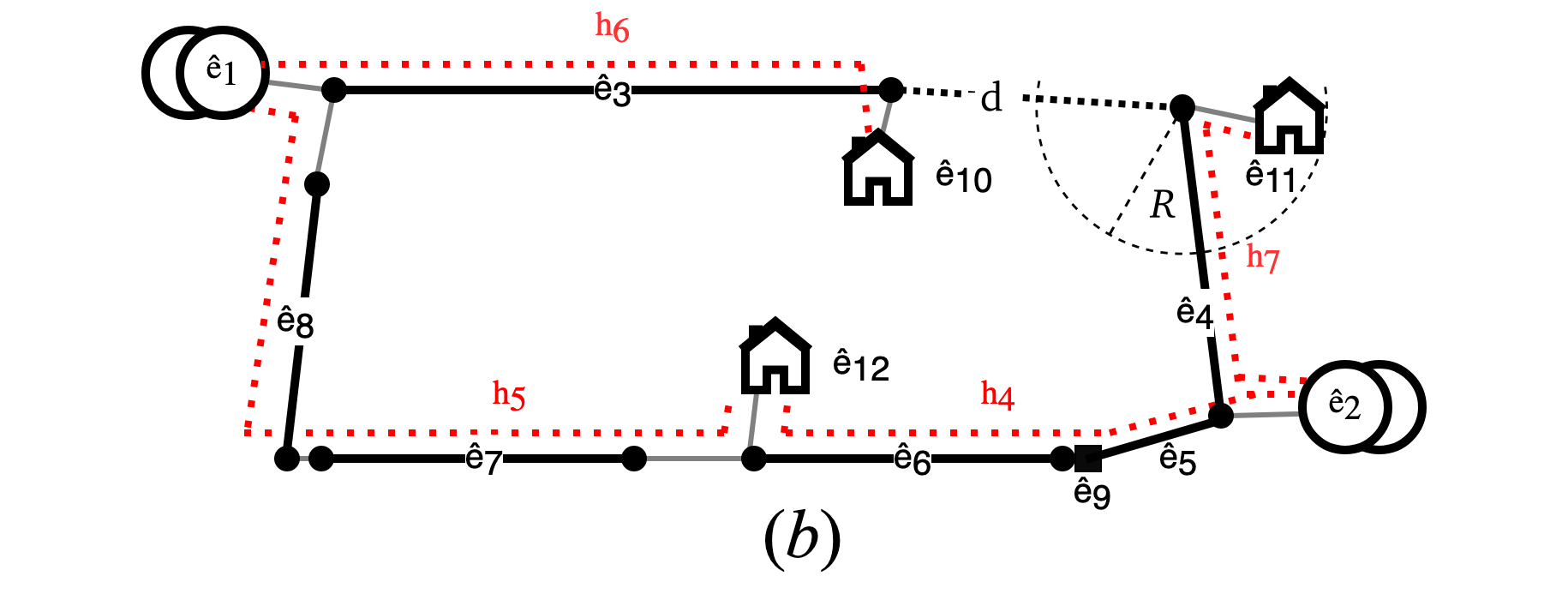}
\includegraphics[width=0.45\textwidth]{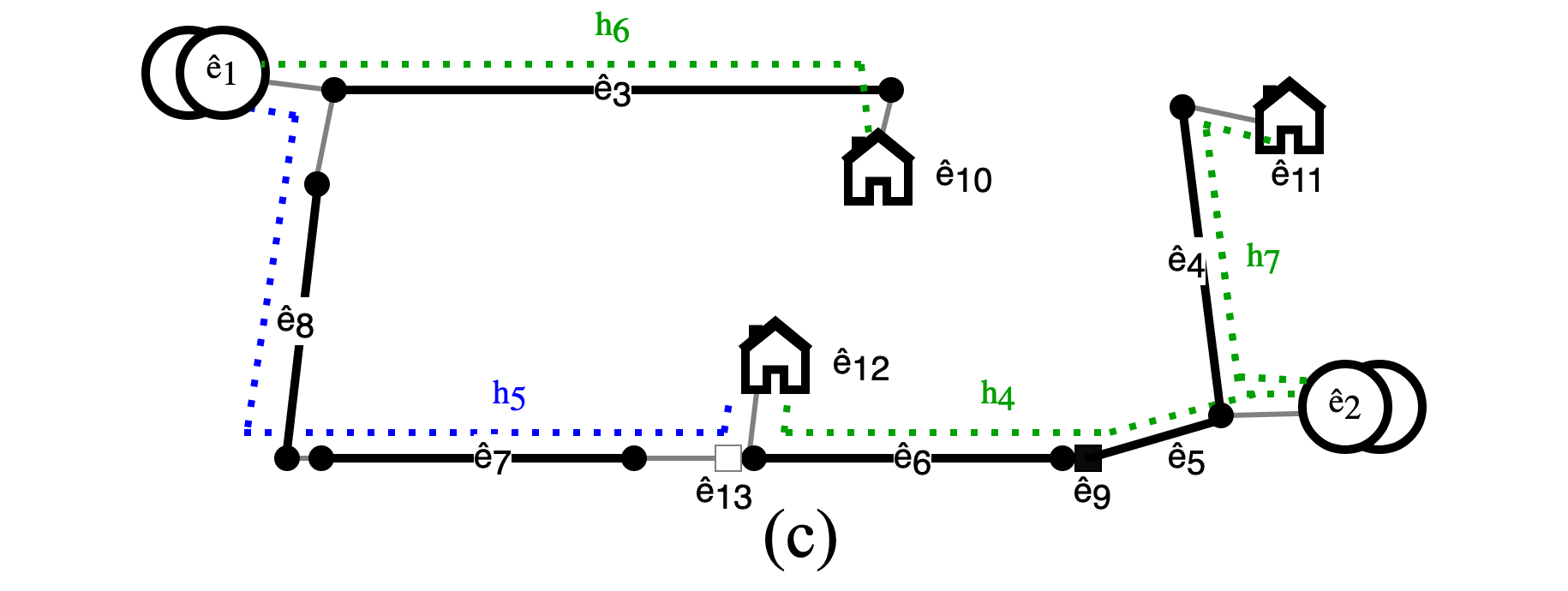}
    
\caption{Application of the systematic procedure to an academic power distribution network. Some hypothetical paths are shown in Fig. \ref{fig:example}a. Figure \ref{fig:example}b shows the network situation after excluding the paths not compatible with the well-defined information from the hypothetical path set. Figure \ref{fig:example}c shows the network situation after modifying the transformation functions, $\mathcal{F}$, to take into account some identified issues. After finding a solution to the TPI problem, it is possible to identify the active and backup paths, as illustrated in Fig. \ref{fig:example}c. }
    \label{fig:example}
\end{figure}

% \subsection{Hypothetical paths}
% Data inaccuracies and missing values underline the complexity of identifying the customer topological paths.
% Inaccurate data and missing elements can result in a disconnected network model, as illustrated in Fig. \ref{fig:example}a.
% Indeed, in such model, there are many possibilities for connecting the network elements causing the existence of a large hypothetical path set, $\mathcal{H}$. \\

\subsection{Hypothetical paths compatible with the well-defined information}
The exact number of hypothetical paths, represented by the set $\mathcal{H}$, cannot be determined without knowing the set of elements $\mathcal{E}$ present in the power distribution network. However, the size of a subset of $\mathcal{H}$, denoted as $\mathcal{H}^{\, i'_1}$, representing the hypothetical paths compatible with the well-defined information $i'_1$ can be determined.\\
To calculate the size of the set $\mathcal{H}^{\, i'_1}$ we need the set of elements $\hat{\mathcal{E}}$ and the set of types $\hat{\mathcal{T}}$. These two sets are obtained using the well-defined information $i'_1$.
Therefore, in this scenario, considering the set of elements $\hat{\mathcal{E}}$ and the set of attributes $\hat{\mathcal{T}}$, the size of the hypothetical paths set, $\mathcal{H}^{\, i'_1}$, is calculated using Eq. (\ref{eq:h}) with $\mathtt{E}=12$, $\mathtt{T}=2$, $\mathtt{C}=3$ and $\mathtt{R}=7$.
\begin{align}   
    |\mathcal{H}^{\, i'_1}| = 82200.
\end{align}

\noindent Some hypothetical paths for certain customers are shown in Fig. \ref{fig:example}a. For example, the hypothetical path $h_1$ composed by the customer $\hat{e}_{12}$ connected to the element $\hat{e}_3$ and the element $\hat{e}_3$ is connected to the transformer $\hat{e}_1$; 
or the hypothetical path $h_3$ composed by the customer $\hat{e}_{11}$ connected to the element $\hat{e}_4$, and following in order: $\hat{e}_5$, $\hat{e}_9$, $\hat{e}_6$, $\hat{e}_7$, $\hat{e}_8$, $\hat{e}_3$ and then the transformer $\hat{e}_1$.\\

\noindent To identify the set of hypothetical paths compatible with the well-defined information, $\mathcal{H}^{\mathcal{I}'}$, the process of reduction is performed as follows. Starting from the set of hypothetical paths $\mathcal{H}^{\, i'_1},$ the process sequentially excludes the paths that are not compatible with the remaining components of the well-defined information, $i'_j \in \mathcal{I}', 1<j<|\mathcal{I}'|$. \\

\noindent Therefore, we determine the set $\mathcal{H}^{\, i'_1, i'_2}$ by excluding the paths $h \in \mathcal{H}^{\, i'_1}$ that are not compatible with the condition of customer's next connection as specified by the well-defined information $i'_2$.\\
For example, the hypothetical path $h_1$, in Fig. \ref{fig:example}a, is excluded since the customer, $\hat{e}_{12}$, next connection is the element $\hat{e}_6$, and not the element $\hat{e}_3$. The size of the set $\mathcal{H}^{\, i'_1, i'_2}$ is given by:
\begin{align}
    |\mathcal{H}^{\, i'_1, i'_2}| = 11742.
\end{align}
\noindent Following, we determine the set $\mathcal{H}^{\, i'_1, i'_2, i'_3}$ by excluding paths $h \in \mathcal{H}^{\, i'_1, i'_2}$ that are not compatible with the condition of proximity as specified by the well-defined information $i'_3$.\\
% in which two consecutive elements, $h_i$ and $h_{i+1}$ with $i<|h|-1$, are not the closest, or their distance is greater than the specified maximum value $R$. Therefore, paths are excluded if the condition $h_{i+1} = Closest(\mathcal{E}, h_i)\text{ AND }Dist(h_i, h_{i+1})<R$ is not satisfied.\\
For example, the path $h_2$, in Fig. \ref{fig:example}a, is excluded since the distance between the elements $\hat{e}_3$ and $\hat{e}_4$ is greater than $R$. The value of $R$, which represents the maximum distance for the connection between two elements, is set to $20$ meters. In this case, the value of $R=20$ meters is chosen based on the assumption that electrical connections between elements in the network are typically within close proximity.
\begin{align}
    |\mathcal{H}^{\, i'_1, i'_2, i'_3}| = 6
\end{align}
\noindent Finally, we determine the set $\mathcal{H}^{\, i'_1, i'_2, i'_3, i'_4}$ excluding the paths, $h \in \mathcal{H}^{\, i'_1, i'_2, i'_3}$ that are not compatible with the condition of length of path as specified by the well-defined information $i'_4$.\\
% whose total length, $l=LengthPath(h)$, is greater than the maximum length $L$, are excluded.\\
For example, the path $h_3$, in Fig. \ref{fig:example}a, is excluded because the length of the path is greater than the $L$. The value of $L$, which to represent the maximum length of a path, is set to $400$ meters. The value $L=400$ meters is chosen to ensure that selected paths are relatively direct and minimise energy loss due to excessive total lengths.
\begin{align}
    |\mathcal{H}^{\, i'_1, i'_2, i'_3, i'_4}| = 4
\end{align}
% So after the reduction, the set of compatible paths for the customer $e_{12}$ is given by two elements: 
% \[
%     \hat{\mathcal{H}}^{\mathcal{I}'} = \{h_2, h_3\}
% \]
\noindent After the process, the set of hypothetical paths is reduced to the paths compatible with the well-defined information. Therefore, the set of hypothetical paths compatible with the well-defined information for the different customers is given by: 
\begin{align}
\mathcal{H}^{\mathcal{I}'} &= \mathcal{H}_{\hat{e}_{10}}^{\mathcal{I}'} \cup \mathcal{H}_{\hat{e}_{11}}^{\mathcal{I}'} \cup \mathcal{H}_{\hat{e}_{12}}^{\mathcal{I}'}, \text{ with}\\
\mathcal{H}_{\hat{e}_{10}}^{\mathcal{I}'} &= \{h_6\}\\
\mathcal{H}_{\hat{e}_{11}}^{\mathcal{I}'} &= \{h_7\} \\
\label{eq:h12}
\mathcal{H}_{\hat{e}_{12}}^{\mathcal{I}'} &= \{h_4, h_5\}.
\end{align}

\subsection{Diagnostic function}
The found solution, $\mathcal{H}^{\mathcal{I}'}$, is evaluated with a $Diagnostic()$ function. This function can process both raw and well-defined information, and it assesses the set $\mathcal{H}^{\mathcal{I}'}$ for any potential issues that may be present.

\noindent One possible validation involves checking the number of active paths for each element of type $customer$ as specified by the raw information $i_5$. This validation is performed by examining the paths for each customer and taking into account the diverse combinations of switch(es) status. This involves examining all possible combinations of switch(es) status and verifying that there is only one active path for each customer, regardless of the switch(es) configuration(s).\\
In this case, during the analysis of the identified paths, it is observed that there are multiple active paths for some customers, as is possible to see in the set $\mathcal{H}_{\hat{e}_{12}}^{\mathcal{I}'}$ in Eq. (\ref{eq:h12}).\\
To address this issue, one solution is to collect more data, for example with a technical meeting with the DSO. 
Therefore, after the meeting with the DSO, a new important piece of information is collected and added to the set of information. This new piece of information is expressed as follows.
\begin{enumerate}
    \item [$i_6$]: one element of type $switch$ is needed for the customer $\hat{e}_{12}$ to disconnect it from the element $\hat{e}_1$.
\end{enumerate}
A new transformation function, $f_5$, is added to process the new information, such that the new well-defined information $i'_5$ is given as follows.
\begin{enumerate}
   \item[$i'_5$]$ = f_5(i_6)$: a new element, $\hat{e}_{13}$, of type $switch$ is added to the set of elements $\hat{\mathcal{E}}$. The location of this new element is chosen to be between the element $\hat{e}_6$ and $\hat{e}_7$ with status open. In this way, the customer $\hat{e}_{12}$ has only one active path. 
   % \item[$i'_6$]$ = f_5(i_5)$: Elements of type $switch$ are added to the set of elements $\hat{mathcal{E}}$ such that no element of type $customer$ can be supplied by more than one element of type $transformer$. 
\end{enumerate}
Given the new well-defined information $i'_5$, the entire procedure is executed one more time, resulting in a new approximation of hypothetical paths, $\mathcal{H}^{\mathcal{I}'}$ is found. 
Following the modification of the transformation functions, the solution is re-evaluated with the $Diagnostic()$ function: the solution is found acceptable, and there is no more information available, so the algorithm is stopped. \\
The set $\mathcal{H}^{\mathcal{I}'}$ is considered the best possible approximation of the set of real paths $\mathcal{P}$, given the information available to the DSO. The approximation found denoted as $\hat{\mathcal{P}}$ is considered the solution for the TPI problem.
% Given all this information, the algorithm is run to find the set of hypothetical paths compatible with the well-defined information, $\mathcal{H}^{\mathcal{I}'}$, that raises no issues when evaluated with the $Diagnostic()$ function.\\
\subsection{Active and backup paths}
Once the solution for the TPI problem is identified, denoted as $\hat{\mathcal{P}}$, it is possible to extract the active and backup paths for each customer. For example, considering the customer $\hat{e}_{12}$ we denote its active and backup path as $\hat{\mathcal{P}}_{a,\hat{e}_{12}}$ and $\hat{\mathcal{P}}_{b, \hat{e}_{12}}$ respectively. These paths are highlighted in Fig. \ref{fig:example}c with the green colour representing the active path and the blue colour representing its possible backup path. More details about the customers paths can be found in Appendix Sections \ref{app:ap} and \ref{app:bp}.
% Note that $\hat{\mathcal{H}}^{\mathcal{I}'}$ is much smaller of $\mathcal{H}$. In this particular example, the size of the set $\hat{\mathcal{H}}^{\mathcal{I}'}$ is $2$ as every customer has an active and backup path.

% \section{Case study} \label{casestudy} 
% \input{Files/CaseStudy}

\section{Scalability Discussion} \label{discussion}
The methodology proposed in Section \ref{methodology} for accurately identifying paths in distribution networks relies on the generation of a large set of hypothetical paths and subsequently excluding the paths that are not compatible with the well-defined information. Even if this approach is effective in identifying the customer paths for small-scale networks, it becomes computationally expensive for larger networks due to the exponential growth in the number of hypothetical paths. The total number of hypothetical paths is given by Eq. \ref{eq:h}. The presence of a factorial term in the equation indicates a rapid growth in the number of hypothetical paths to consider, resulting in an increase of computational cost as the number of network elements increases.\\

% There may be many possible solutions to solve the computation issue such as probabilistic methods, machine learning methods
\noindent To address the scalability challenges, we propose an alternative approach that we call the expanding path set (EPS) technique. This technique focuses on recursively expanding the set of hypothetical paths: starting with a small set of hypothetical paths and iteratively growing the set based on the available well-defined information.
A possible implementation of this alternative approach is detailed by the algorithm given in Appendix Section \ref{sec:aa}.\\

\noindent While both algorithms, the excluding and expanding one, come up with the same solution, the algorithm growing the set is more efficient since only the paths that are compatible with the well-defined information are considered.

\section{Conclusion} \label{conclusion}
% \noindent In this paper, we propose a systematic approach for topological paths identification in low-voltage power distribution networks. The proposed methodology is based on a systematic framework that takes into account the partial data, available information and necessary assumptions. We apply this methodology to an actual distribution network located in Belgium, where we identify paths that are compatible with the provided information and assumptions. This practical application highlights the effectiveness and applicability of our approach.\\
% Such approach can be used to improve the understanding of the network structure, which can be used to plan, to identify potential bottlenecks, and to assess the impact of DERs on the network.
% The proposed approach is a valuable tool for DSOs to manage networks, and it is a significant step forward in the development of methods for topological paths identification in power distribution networks.

In conclusion, this paper introduced a systematic procedure to address the complex challenge of topological path identification (TPI) within power distribution networks.
% Distribution system operators (DSOs) face the complex task of identifying the active and backup paths for customers connected to different transformers.
% a task made more complex by the large amount of raw information available and the high number of network's elements.
By leveraging a methodology that integrates raw information, transformation functions that transform the raw information into well-defined information, and hypothetical paths compatible with the well-defined information, the proposed approach effectively identifies the customer paths within the power distribution network. This systematic procedure bridges the gap between data scientists and network experts, facilitating their collaboration in resolving the TPI problem in a consistent and efficient manner. As a result, this work significantly contributes to enhancing the efficiency and reliability of power distribution network modelling and management.
% The ability for DSOs to identify and manage topological paths brings a multitude of benefits. These include an essential communication channel to inform customers of network maintenance, proactively identifying and mitigating potential power quality issues, and facilitating decision-making in network investment planning.
% We argue our method is more reliable than other existing methods where there is no explicit mention of the transformation applied to the data, so in the case of unexpected outputs, it is not possible to exactly point out whether there were some mistakes in the process or in the data. 
% Despite the fact that our method can be applied to any DSO's situation, we applied this procedure to a real power distribution network in Belgium to demonstrate its effectiveness and applicability.
% Its potential to help DSOs in solving the complexities of their grids suggests a brighter future for power distribution networks, where communication with customers, efficiency, and proactive decision-making are essential.
\\The future works for this paper are to apply our procedure to a real distribution network in Belgium. Additionally, future research could explore methods for identifying more efficient ways to determine parameters such as the maximum distance $R$ and the total path length $L$, potentially considering distributions of values rather than single values, employing suitable probability distribution functions. This would allow for a more systematic and adaptable approach to path identification in real-world scenarios.

\section*{References}
\printbibliography[heading=none]

% \newpage
\appendix
\label{appendix}
\subsection{Real network elements}
Table \ref{tab:rne} provides a detailed overview of all the information regarding the elements in the real network, including their type, coordinates, and status. It is possible to note that some elements of type $line$ can have more than two pairs of coordinates, as in the case of element $e_{13}$.
%\vspace{-4mm}
\begin{table}[h]
\centering
\caption{Real network elements and their attributes}
\label{tab:rne}
\resizebox{0.9\columnwidth}{!}{%
\begin{tabular}{|c|c|c|c|}
\hline
\textbf{ID} & \textbf{Coordinates}                     & \textbf{Type} & \textbf{Status} \\ \hline
$e_1$       & (232, 299)                               & transformer   & -             \\ \hline
$e_2$       & (399, 196)                               & transformer   & -             \\ \hline
$e_3$       & (232, 299), (241, 298)             & line          & -             \\ \hline
$e_4$       & (241, 298), (327, 297)             & line          & -             \\ \hline
$e_5$       & (327, 297), (324, 287)             & line          & -             \\ \hline
$e_6$       & (388, 295), (393, 292)             & line          & -             \\ \hline
$e_7$       & (388, 295), (390, 193)             & line          & -             \\ \hline
$e_8$       & (390, 193), (399, 196)             & line          & -             \\ \hline
$e_9$       & (390, 193), (376, 191)             & line          & -             \\ \hline
$e_{10}$      & (376, 191), (315, 191)             & line          & -             \\ \hline
$e_{11}$      & (315, 191), (295, 189)             & line          & -             \\ \hline
$e_{12}$      & (295, 189), (299, 207)             & line          & -             \\ \hline
$e_{13}$      & (295, 189), (268, 169), (235, 190) & line          & -             \\ \hline
$e_{14}$      & (235, 190), (239, 282)             & line          & -             \\ \hline
$e_{15}$      & (239, 282), (241, 298)             & line          & -             \\ \hline
$e_{16}$      & (376, 191)                               & switch        & open            \\ \hline
$e_{17}$      & (315, 191)                               & switch        & open            \\ \hline
$e_{18}$      & (239, 282)                               & switch        & close           \\ \hline
$e_{19}$      & (324, 287)                               & customer      & -             \\ \hline
$e_{20}$      & (393, 292)                               & customer      & -             \\ \hline
$e_{21}$      & (299, 207)                               & customer      & -             \\ \hline
\end{tabular}%
}
\end{table}
%\vspace{-4mm}

\subsection{DSO network elements}
Table \ref{tab:dne} shows all the details about the elements initially available to the DSO. It is possible to note that elements of type $line$ have only two pairs of coordinates, since intermediate points are generally not known by DSOs.
%\vspace{-4mm}
\begin{table}[h!]
\centering
\caption{Elements available to the DSO and their attributes}
\label{tab:dne}
\resizebox{0.9\columnwidth}{!}{%
\begin{tabular}{|c|c|c|c|}
\hline
\textbf{ID}    & \textbf{Coordinates}            & \textbf{Type}        & \textbf{Status} \\ \hline
$\hat{e}_1$    & (230, 297)             & transformer & -                                \\ \hline
$\hat{e}_2$    & (401, 194)             & transformer & -                                \\ \hline
$\hat{e}_3$    & (240, 297), (328, 296) & line        & -                                \\ \hline
$\hat{e}_4$    & (388, 295), (391, 193) & line        & -                                \\ \hline
$\hat{e}_5$    & (390, 193), (374, 189) & line        & -                                \\ \hline
$\hat{e}_6$    & (374, 189), (310, 191) & line        & -                                \\ \hline
$\hat{e}_7$    & (293, 189), (235, 189) & line        & -                                \\ \hline
$\hat{e}_8$    & (233, 189), (242, 282) & line        & -                                \\ \hline
$\hat{e}_9$    & (374, 189)             & switch      & close                            \\ \hline
$\hat{e}_{10}$ & (321, 289)             & customer    & -                                \\ \hline
$\hat{e}_{11}$ & (393, 292)             & customer    & -                                \\ \hline
$\hat{e}_{12}$ & (320, 196)             & customer    & -                                \\ \hline
\end{tabular}%
}
\end{table}
%\vspace{-4mm}

\subsection{Active paths}
\label{app:ap}
\begin{figure}[h!]
    \centering
\includegraphics[width=0.42\textwidth]{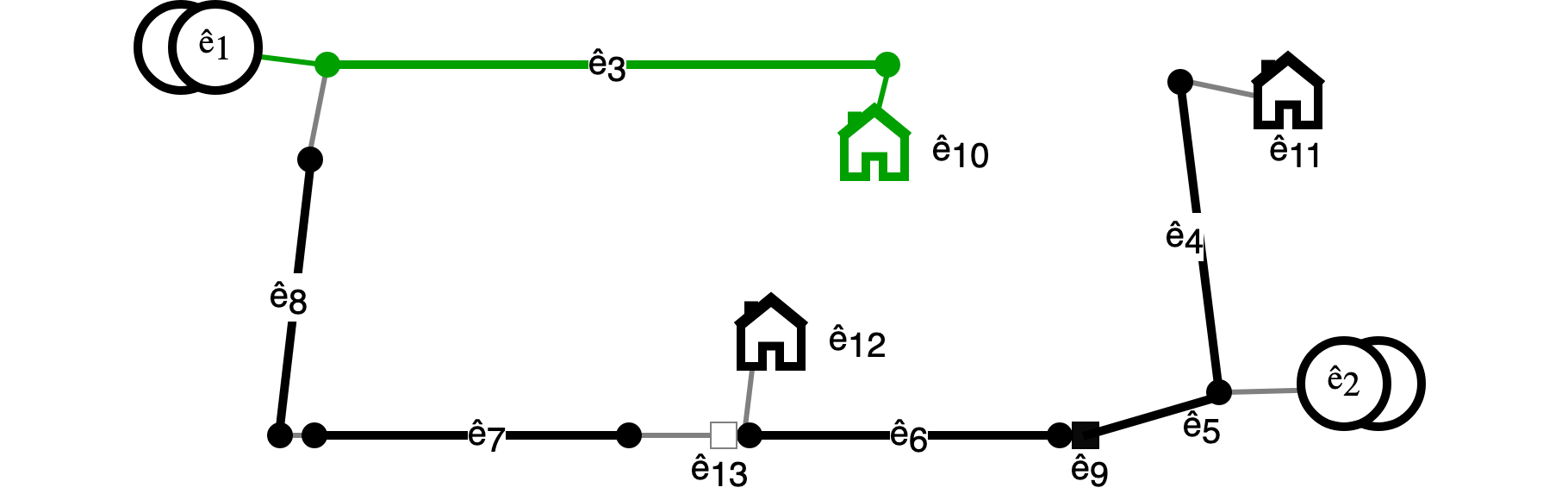}
\caption{Active path for customer $\hat{e}_{10}$}
    \label{fig:c10A}
\end{figure}

\begin{figure}[h!]
    \centering
\includegraphics[width=0.42\textwidth]{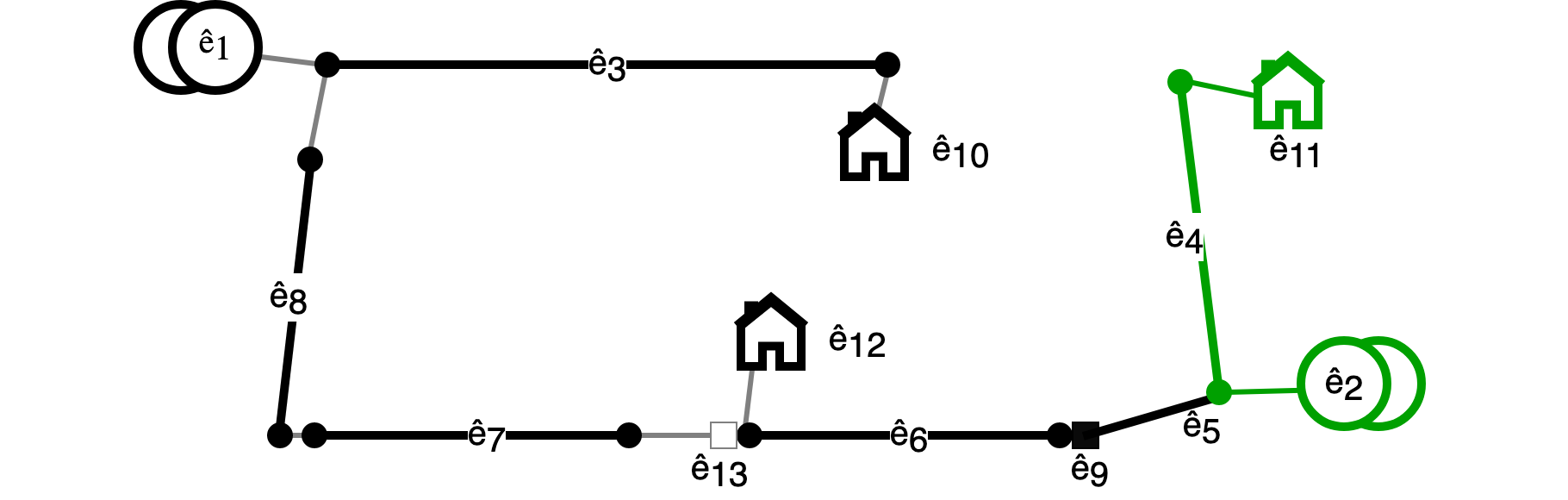}
\caption{Active path for customer $\hat{e}_{11}$}
    \label{fig:c11A}
\end{figure}

\begin{figure}[h!]
    \centering
\includegraphics[width=0.42\textwidth]{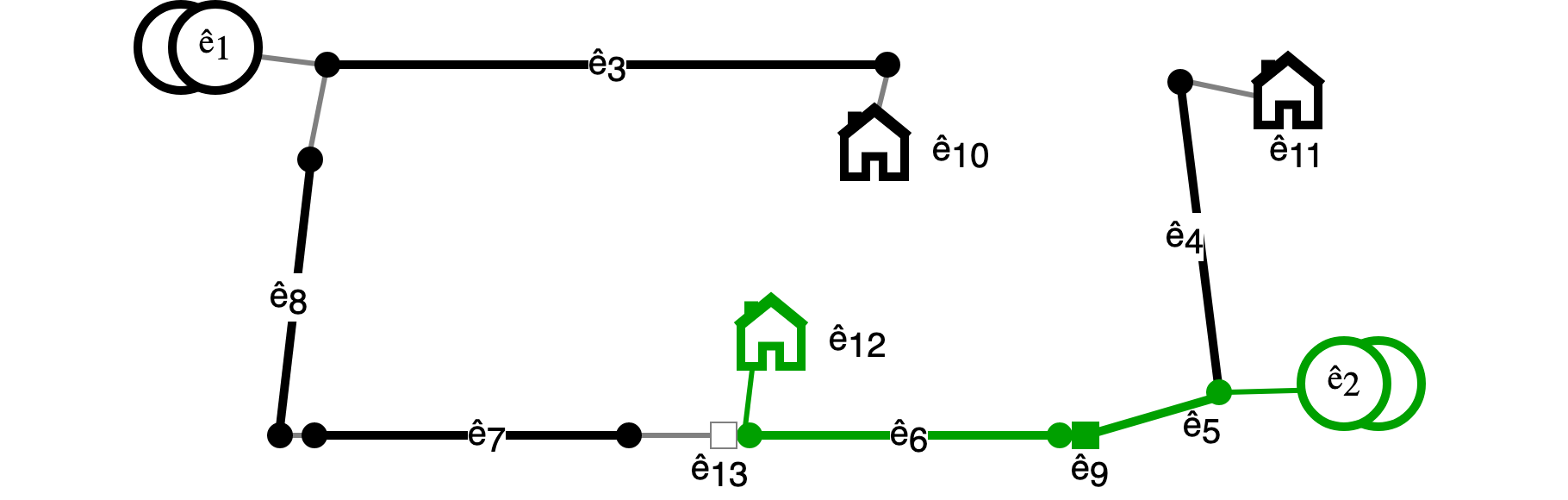}
\caption{Active path for customer $\hat{e}_{12}$}
    \label{fig:c12A}
\end{figure}

Table \ref{tab:cap} shows all the active paths for each customer. Moreover, Fig. \ref{fig:c10A}, \ref{fig:c11A} and \ref{fig:c12A} show the hypothetical active paths compatible with the well-defined information for the customers $\hat{e}_{10}$, $\hat{e}_{11}$ and $\hat{e}_{12}$ respectively.
\begin{table}[h]
\centering
\caption{}
\label{tab:cap}
\resizebox{0.75\columnwidth}{!}{%
\begin{tabular}{|c|c|}
\hline
\textbf{Customer ID} & \textbf{Paths}                             \\ \hline
$\hat{e}_{10}$       & ($\hat{e}_{10}$, $\hat{e}_3$, $\hat{e}_1$) \\ \hline
$\hat{e}_{11}$       & ($\hat{e}_{11}$, $\hat{e}_4$, $\hat{e}_2$) \\ \hline
$\hat{e}_{12}$ & ($\hat{e}_{12}$, $\hat{e}_6$, $\hat{e}_9$, $\hat{e}_5$, $\hat{e}_4$, $\hat{e}_2$) \\ \hline
\end{tabular}%
}
\end{table}


\subsection{Backup paths}
\label{app:bp}
\begin{figure}[h!]
    \centering
\includegraphics[width=0.42\textwidth]{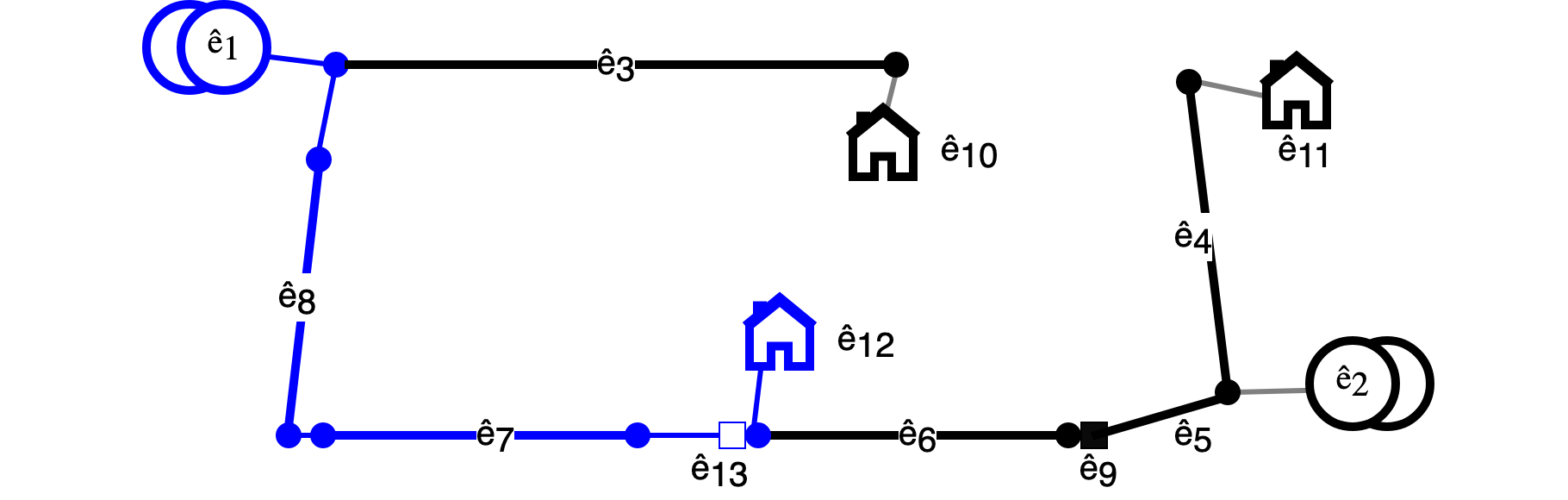}
    
\caption{Backup path for customer $\hat{e}_{12}$}
    \label{fig:c12B}
\end{figure}

Table \ref{tab:cab} showing all the backup paths for each customer. Fig. \ref{fig:c12B} shows the hypothetical backup paths compatible with the well-defined information for the customer $\hat{e}_{12}$.\\
%\vspace{-3mm}
\begin{table}[h!]
\centering
\caption{}
\label{tab:cab}
\resizebox{0.75\columnwidth}{!}{%
\begin{tabular}{|c|c|}
\hline
\textbf{Customer ID} & \textbf{Paths}                             \\ \hline
$\hat{e}_{10}$       & None \\ \hline
$\hat{e}_{11}$       & None \\ \hline
$\hat{e}_{12}$       & ($\hat{e}_{12}$, $\hat{e}_6$, $\hat{e}_{13}$, $\hat{e}_7$, $\hat{e}_8$, $\hat{e}_3$, $\hat{e}_1$) \\ \hline
\end{tabular}%
}
\end{table}

\subsection{Expansion path set algorithm}
\label{sec:aa}

\algrenewcommand\algorithmicindent{0.42em} %change identation

\noindent The expansion path set (EPS) algorithm works as follows. For each customer, the $IdentifyPaths()$ function initialises an empty path and iteratively explores the paths considering adjacent elements. Paths that fail to meet the constraints specified by the well-defined information of the academic example explained in Section \ref{accexample} (distance between elements, total path length limits) are not further expanded. If an element of type $transformer$ is reached, the algorithm is stopped, and the path is returned. Otherwise, the algorithm recursively explores new paths invoking the same function with the updated parameters. \\
The recursiveness of the algorithm operates similar to an expansion tree: as the algorithm examines each element, it branches out to explore all possible connections that lead to paths that are compatible with the set of well-defined information.\\

\begin{algorithm}[h]
\textbf{Function:} $IdentifyPaths(current\_element, current\_path, R, L)$
\hspace*{\algorithmicindent} \textbf{Inputs:}
\begin{itemize}[leftmargin=4em]
    \item $current\_element$: the element to consider as a starting point to find its path(s) to the transformer(s).
    \item $current\_path$: the current path being considered.
    \item $R$: the maximum distance criterion for path compatibility.
    \item $L$: the maximum path length criterion for path compatibility.
\end{itemize}
\hspace*{\algorithmicindent} \textbf{Output:}
\begin{itemize}[leftmargin=4em]
    \item $candidate\_paths$: list of paths compatible with the set of well-defined information.
\end{itemize}

\begin{algorithmic}[1]
  \State $candidate\_paths$ = []
  \ForAll{$next\_element$ in $\big(\hat{\mathcal{E}} - Subset(\hat{\mathcal{E}}, customer)\big)$}
    \If{$next\_element$ not in $current\_path$}
      \State $distance = Dist(current\_element, next\_element)$
      \State $new\_path = current\_path + [next\_element]$
      \State $length = LengthPath(new\_path)$
      \If{$(distance < R$ and $ length < L )$}
        \If{$next\_element.type=transformer$}
          \State $candidate\_paths.add(new\_path)$
        \Else
            \State $paths = IdentifyPaths(next\_element, new\_path, R, L)$
            \State $candidate\_paths.add(paths)$
        \EndIf
      \EndIf
    \EndIf
  \EndFor
  \State \Return $candidate\_paths$
% \EndFunction
\end{algorithmic}
\end{algorithm}

\end{document}